\documentclass[superscriptaddress,aps,prl,reprint,twocolumn,amsmath,amssymb,showpacs,]{revtex4-2}
\usepackage{graphicx}  
\usepackage{newtxtext, newtxmath}
\usepackage{epstopdf}
\usepackage{dcolumn}   
\usepackage{bm}        
\usepackage{amssymb}   
\usepackage{amsmath}   

\usepackage{amsthm}
\usepackage{chemarrow}
\usepackage{color}
\usepackage{mathrsfs}
\usepackage{float}
\usepackage{bbold}
\usepackage{bbm}
\usepackage{braket}
\usepackage{physics}
\usepackage[normalem]{ulem}
\usepackage{upgreek}

\usepackage[a4paper,colorlinks=true,
linkcolor=blue,citecolor=blue,urlcolor=black,
pdfauthor={ },
pdftitle={ },
pdfsubject={ },
pdfkeywords={ }]{hyperref}

\theoremstyle{plain}
\newtheorem*{theorem*}{Theorem}

\usepackage{verbatim}

\usepackage{xcolor}

\begin{document}


\title{Hyperpolarized Molecular Nuclear Spins Achieve Magnetic Amplification} 

\affiliation{
Laboratory of Spin Magnetic Resonance, School of Physical Sciences, Anhui Province Key Laboratory of Scientific Instrument Development and Application, University of Science and Technology of China, Hefei 230026, China}
\affiliation{
\mbox{Hefei National Laboratory, University of Science and Technology of China, Hefei 230088, China}}
\affiliation{
Hefei National Research Center for Physical Sciences at the Microscale, University of Science and Technology of China, Hefei 230026, China}

 \author{Shengbang Zhou}
 \affiliation{
 Laboratory of Spin Magnetic Resonance, School of Physical Sciences, Anhui Province Key Laboratory of Scientific Instrument Development and Application, University of Science and Technology of China, Hefei 230026, China}
 \affiliation{
 Hefei National Research Center for Physical Sciences at the Microscale, University of Science and Technology of China, Hefei 230026, China}

 \author{Qing Li}
 \affiliation{
 Laboratory of Spin Magnetic Resonance, School of Physical Sciences, Anhui Province Key Laboratory of Scientific Instrument Development and Application, University of Science and Technology of China, Hefei 230026, China}
 \affiliation{
 Hefei National Research Center for Physical Sciences at the Microscale, University of Science and Technology of China, Hefei 230026, China}

  \author{Yi Ren}
 \affiliation{
 Laboratory of Spin Magnetic Resonance, School of Physical Sciences, Anhui Province Key Laboratory of Scientific Instrument Development and Application, University of Science and Technology of China, Hefei 230026, China}

  \author{Jingyan Xu}
 \affiliation{
 Institute of Physics, Johannes Gutenberg University of Mainz, 55099 Mainz, Germany}
 \affiliation{
 Helmholtz Institute Mainz, 55099 Mainz, Germany}
 \affiliation{
 GSI Helmholtzzentrum fur Schwerionenforschung, 64291 Darmstadt, Germany}

  \author{Raphael Kircher}
 \affiliation{
 Institute of Physics, Johannes Gutenberg University of Mainz, 55099 Mainz, Germany}
 \affiliation{
 Helmholtz Institute Mainz, 55099 Mainz, Germany}
 \affiliation{
 GSI Helmholtzzentrum fur Schwerionenforschung, 64291 Darmstadt, Germany}

 \author{Danila A. Barskiy}
\affiliation{
 Institute of Physics, Johannes Gutenberg University of Mainz, 55099 Mainz, Germany}
 \affiliation{
 Helmholtz Institute Mainz, 55099 Mainz, Germany}
 \affiliation{
 GSI Helmholtzzentrum fur Schwerionenforschung, 64291 Darmstadt, Germany}
 \affiliation{Frost Institute for Chemistry and Molecular Science, Department of Chemistry, University of Miami, Coral Gables, FL 33146, USA}

  \author{Dmitry Budker}
 \affiliation{
 Institute of Physics, Johannes Gutenberg University of Mainz, 55099 Mainz, Germany}
 \affiliation{
 Helmholtz Institute Mainz, 55099 Mainz, Germany}
 \affiliation{
 GSI Helmholtzzentrum fur Schwerionenforschung, 64291 Darmstadt, Germany}
 \affiliation{
 Department of Physics, University of California, Berkeley, CA 94720, USA}

 \author{Min Jiang}
 \email[]{dxjm@ustc.edu.cn}
 \affiliation{
 Laboratory of Spin Magnetic Resonance, School of Physical Sciences, Anhui Province Key Laboratory of Scientific Instrument Development and Application, University of Science and Technology of China, Hefei 230026, China}
 \affiliation{
 \mbox{Hefei National Laboratory, University of Science and Technology of China, Hefei 230088, China}}
 \affiliation{
 Hefei National Research Center for Physical Sciences at the Microscale, University of Science and Technology of China, Hefei 230026, China}

 \author{Xinhua Peng}
 \email[]{xhpeng@ustc.edu.cn}
 \affiliation{
 Laboratory of Spin Magnetic Resonance, School of Physical Sciences, Anhui Province Key Laboratory of Scientific Instrument Development and Application, University of Science and Technology of China, Hefei 230026, China}
 \affiliation{
 \mbox{Hefei National Laboratory, University of Science and Technology of China, Hefei 230088, China}}
 \affiliation{
 Hefei National Research Center for Physical Sciences at the Microscale, University of Science and Technology of China, Hefei 230026, China}

\date{\today}


\begin{abstract}

The use of nuclear spins as physical sensing systems is disadvantaged by their low signal responsivity, particularly when compared to sensing techniques based on electron spins.
This primarily results from the small nuclear gyromagnetic ratio and the difficulties in achieving high spin polarization.
Here we develop a new approach to investigating the response of hyperpolarized molecular nuclear spins to magnetic fields
and demonstrate orders-of-magnitude enhanced magnetic responsivity
over state-of-the-art proton and Overhauser magnetometers.
Using hyperpolarized molecules with proton spins,
we report the realization of magnetic amplification in linear and nonlinear types.
We further extend this amplification to hyperpolarized scalar-coupled multi-spin molecules and observe substantial magnetic amplification exceeding $10\%$.
Moreover, we observe an anomalous amplification with dispersive frequency dependence that originates from magnetic interference effects.
Our work highlights the potential of hyperpolarized molecular nuclear spins for use in a new class of quantum sensors, with promising applications in both applied and fundamental physics, including highly accurate absolute magnetometry and the exploration of axion–nucleon exotic interactions.

\end{abstract}

\maketitle

Nuclear spins,
known for their long coherence times under ambient conditions and ubiquitous presence in matter,
have enabled widespread applications across chemical analysis, medical imaging, and materials science\,\cite{1990principles,1984Medical, 2008spindynamics,2009TScience}.
Despite their successful applications in these fields, nuclear spins have not fully showcased their significant advantages in quantum sensing as physical sensing systems.
For example,
while various well-established nuclear-spin sensors,
such as proton magnetometers\,\cite{2021proton,gross2016dynamic,ledbetter2012liquid,wu2018nuclear} and Overhauser magnetometers\,\cite{2017Overhauser, 2025Overhauser,2021Overhauser}, are widely used,
they continue to face challenges related to limited sensitivity.
This is primarily due to their small nuclear gyromagnetic ratio and low polarization of nuclear spins, especially when compared to electron-spin based sensors\,\cite{2003mag, 2007mag, 2018mag, 2021NV}.
Recently, the demand for developing ultrasensitive sensing has increased extensively in areas such as absolute magnetometry\,\cite{2021geophysic, 20203He,2009geomagnetic}, gyroscope technology\,\cite{2005Gyroscope, 2021Gyroscope}, and fundamental physics tests\,\cite{2014CASPEr, 2018axion, 2019natureaxion, 2021NPaxion, 2019axion,jiang2024searches,cong2025spin}.
These emphasize
the unique measurements of nuclear-dependent interactions
necessitating advancement of sensing techniques based on nuclear spins.
Recent progress in the hyperpolarization methods of nuclear spins in various molecules,
including parahydrogen-induced polarization (PHIP)\,\cite{2023hyperreviews, 1991PHIP, amp_NP2009science, 2011PHIP} and dynamic nuclear polarization (DNP)\,\cite{2003DNPPNAS, 2021DNP},
provides a promising path to overcoming the low nuclear polarization barrier, potentially increasing polarization by several orders of magnitude.
To date, these hyperpolarization methods, predominantly applied in NMR spectroscopy\,\cite{2023hyperreviews, 1991PHIP, amp_NP2009science, 2011PHIP,2003DNPPNAS, 2021DNP},
have seen limited exploration within the field of metrology.
The potential combination of nuclear-spin sensing with hyperpolarization methods may open up exciting opportunities across a wide range of applications\,\cite{2021geophysic, 20203He,2009geomagnetic,2005Gyroscope, 2021Gyroscope,2014CASPEr, 2018axion, 2019natureaxion, 2021NPaxion, 2019axion,jiang2024searches,cong2025spin}.

To evaluate the potential of hyperpolarized molecular nuclear spins as sensors,
we first provide an approach to analyzing their signal responsivity.
Generally, the physical principle of spin-based sensing involves an ensemble of polarized nuclei which, when exposed to a measured oscillating magnetic field with amplitude $B_\text{a}$,
generate a dipolar magnetic field with an amplitude $B_\text{s}$.
This induced dipolar field can be further detected by external detectors such as pick-up coils\,\cite{1990principles}, atomic magnetometers\,\cite{2003mag, 2007mag, 2018mag}, and SQUIDs\,\cite{1998squid, 2002squid}.
In this work, we focus solely on the intrinsic responsivity of nuclear spins,
excluding the influence of external detectors.
We define an amplification factor ${G = B_\text{s} / B_\text{a}}$ that can quantify the spin-based sensing performance.
As derived in the Supplemental Material\,\cite{supp}, this factor can be further determined by
\begin {equation}
G = 2 \xi\mu_0 T_2 \gamma V M_0 = \xi\mu_0 T_2\gamma^2\hbar n V P_0\,,
\label{G}
\end {equation}
where $\mu_0$ is the permeability of free space,
$T_2$ is the nuclear spin transverse relaxation time,
$\gamma$ is the nuclear gyromagnetic ratio,
$V$ is the volume of sensing molecules,
$M_0$ is the initial magnetization,
and $\xi$ is a factor accounting for geometric considerations and detection distance.
The magnetization $M_0$ for spin-1/2 particles is given by ${M_0 = \left(1/2\right)\hbar\gamma n P_0}$,
where $\hbar$ is Planck’s constant,
and $n$ and $P_0$ are the spin number density and the degree of nuclear polarization, respectively.
Equation\,\ref{G} provides a useful approach to evaluate the performance of both existing and potential nuclear-spin sensors.
For example, in an Overhauser magnetometer using protons as the sensing system\,\cite{2017Overhauser, 2025Overhauser, 2021Overhauser}, 
with ${\gamma_\text{H} \approx 2\pi \times 42.576 \, \text{MHz/T}}$, ${T_2 \approx 2 \, \text{s}}$,
and a total number of polarized proton spins ${nVP_0 \approx 10^{15}}$,
the corresponding amplification factor is estimated to be $G \approx 0.001\%$.
In contrast, recent experiments in parahydrogen-enhanced NMR spectroscopy demonstrate that $n V P_0$ can reach values on the order of $10^{17}$\,\cite{amp_NP2009science, amp_NP2015, amp_NP2016PNAS, amp_NP2019, 2015SabrePy}.
Such quantities of polarized spins could yield a potential amplification factor of ${G \approx 0.1\%}$,
representing a potential improvement of two orders of magnitude.
Despite the promise of hyperpolarized molecular nuclear spins for precision metrology, significant experimental effort is still required to fully realize this potential.


In this Letter, we present the first demonstration of magnetic amplification using hyperpolarized molecular nuclear spins.
The central ingredient involves preparing hyperpolarized molecular nuclear spins of neat liquids characterized by a high number of polarized spins and long-lived nuclear coherence,
achieved through the PHIP method.
Using molecules including acetonitrile and pyridine, we demonstrate that with hyperpolarized protons,
one can achieve magnetic field amplification factors up to ${G \approx 3.1\%}$.
This approach is further extended to scalar-coupled multi-spin molecules, reaching amplification factors of ${G \approx 13.2\%}$, attributed to their advantageous long-lived states at zero field.
Our results show an improvement of at least three orders of magnitude over state-of-the-art proton and Overhauser magnetometers\,\cite{2021proton,gross2016dynamic,ledbetter2012liquid,wu2018nuclear,2017Overhauser, 2025Overhauser,2021Overhauser}.
We also observe an anomalous frequency dependence of the amplification, and demonstrate that this is generated by the interference between the oscillating field and the induced dipolar field.
Our work offers significant potential for both applied and fundamental physics, including applications in highly accurate absolute magnetometry\,\cite{2021geophysic, 20203He,2005Gyroscope, 2021Gyroscope,2009geomagnetic} and the searches for axion–nucleon exotic interactions\,\cite{2014CASPEr, 2018axion, 2019natureaxion, 2021NPaxion, 2019axion}.



\begin{figure}[t]  
	\makeatletter
	\def\@captype{figure}
	\makeatother
	\includegraphics[scale=0.86]{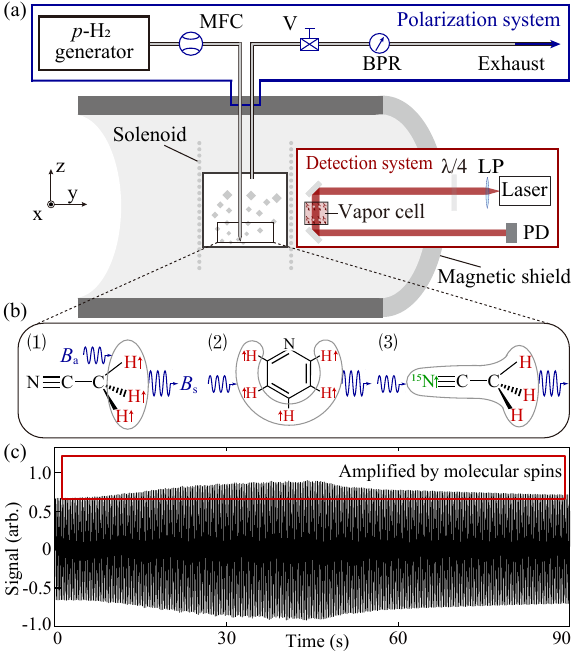}
	\caption{Magnetic amplification using hyperpolarized molecular nuclear spins. (a) Setup. The polarization system is designed to polarize molecular nuclear spins by bubbling $p$-H$_2$ gas through a neat liquid, and consists of a parahydrogen ($p$-$\text{H}_2$) generator, mass flow controller (MFC), valve (V) and back pressure regulator (BPR).
    The dipolar fields generated by the resulting hyperpolarized nuclear spins are detected using a $^{87}$Rb magnetometer, which includes a laser beam passed through a linear polarizer (LP), a quarter-wave plate ($\uplambda$/4) and a vapor cell, detected with a photodiode (PD)\,\cite{supp}. (b) Schematic of magnetic amplification using proton spins from acetonitrile (1) and pyridine (2), as well as scalar-coupled spins from $^{15}$N-labeled acetonitrile (3).
    (c) Amplification signal using $^{15}$N-labeled acetonitrile.
    }
	\label{fig1}
\end{figure}

Our experiments are conducted using the setup depicted in Fig.\,\ref{fig1}(a).
Approximately ${0.5}$\,mL of the liquid of sensing molecules along with an iridium-based catalyst is contained within a glass tube [Fig.\,\ref{fig1}(a)].
Hyperpolarized molecular nuclear spins are prepared using a non-hydrogenative PHIP method known as signal amplification by reversible exchange (SABRE)\,\cite{amp_NP2009science, 2015SabrePy, 2019SABREreviews,2025benchtop},
where parahydrogen ($p$-$\text{H}_2$) gas is continuously bubbled into the glass tube.
Additional details on polarization are provided in the Supplemental Material\,\cite{supp}.
Once polarization is complete, a measured field $B_\text{a}$ is applied to the hyperpolarized spins, and they generate a dipolar field $B_\text{s}$, which is detected using an atomic magnetometer\,[Fig.\ref{fig1}(a)]\,\cite{supp}.
In experiments, we use proton spins from acetonitrile and pyridine, and scalar-coupled spins from $^{15}$N-labeled acetonitrile as physical sensors [Fig.\,\ref{fig1}(b)].
A typical time-domain signal, obtained using $^{15}$N-labeled acetonitrile as sensing molecules, is shown in Fig.\,\ref{fig1}(c).
The data clearly show that the dipolar field $B_\text{s}$ generated by the hyperpolarized molecular nuclear spins amplifies the signal amplitude.

Maximizing the amplification factor $G$ requires achieving both high spin density $n$ and polarization $P_0$, according to Eq.\,\ref{G}.
We use two types of sensing molecules, pyridine and acetonitrile, each with an iridium-based catalyst, for magnetic sensing.
Unlike previous studies that used diluted solutions for chemical analysis\,\cite{amp_NP2009science, 2015SabrePy, 2019SABREreviews,2025benchtop},
we use neat liquid to greatly increase the spin density to approximately ${n\approx 10^{22}\,\text{cm}^{-3}}$.
However, using neat liquid imposes a limitation on achieving high spin polarization compared to diluted solutions.
For protons in acetonitrile or pyridine,
this polarization is experimentally calibrated to be $P_0\approx 0.02\%$,
which corresponds to the thermal spin polarization at 300\,K under about $60$\,T.
Fortunately, it is the number of polarized spins that determines the amplification, rather than the spin polarization alone.
Overall,
we achieve a substantial number of polarized molecular nuclear spins, approximately $nVP_0 \approx  10^{18}$\,\cite{supp}.

\begin{figure}[t]  
	\makeatletter
	\def\@captype{figure}
	\makeatother
	\includegraphics[scale = 0.96]{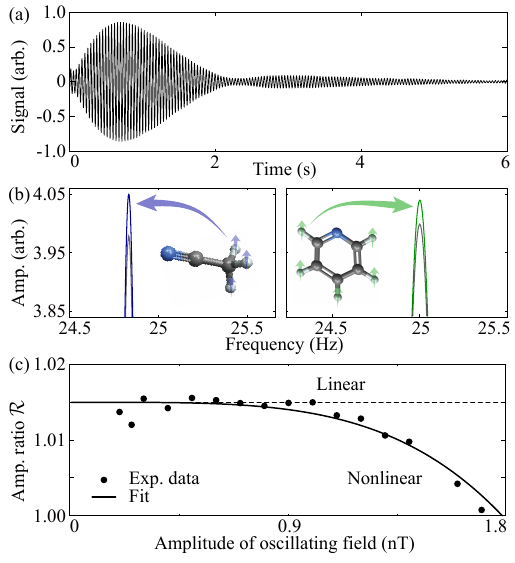}
	\caption{Demonstrations of magnetic amplification using proton spins. (a) Real-time response of proton spins to a near-resonant oscillating field. (b) Magnetic amplification using the hyperpolarized protons of acetonitrile (left) and pyridine (right).
    Colored curves show the spectral amplitude in the presence of hyperpolarized protons, while gray curves correspond to the oscillating field alone.
    The complete spectra are provided in the Supplemental Material\,\cite{supp}.
    (c) Dependence of the ratio $\mathcal{R}$ on the amplitude of oscillating field.}
	\label{fig2}
\end{figure}


We begin by examining the magnetic responsivity of hyperpolarized molecular nuclear spins.
Our analysis starts with a simple two-level system, such as proton spins, and is then expanded to scalar-coupled spins.
Unlike the idealized scenario in Eq.\,\ref{G},
where spin polarization is assumed to remain constant and the measured field is perfectly resonant with sensing spins,
our following analysis incorporates the effects of spin decoherence and the frequency-dependent behavior of hyperpolarized nuclear spins.
At time $t = 0$, the spins are prepared along the $z$-axis bias field $B_0$.
As the system evolves, the amplification factor (i.e., ${G = \partial B_\text{s} / \partial B_\text{a}}$) is modified to\,\cite{supp}
\begin{equation}
G = \xi \mu_0 \gamma^2 \hbar n V P_0 t e^{-t/T_2} \text{sinc}(\Delta t/2)\,,
\label{eq.BsoverBa}
\end{equation}
where $t$ is the sensing time,
$\Delta = 2\pi(\nu_\text{a}-\nu_0)$ represents the frequency detuning,
$\nu_\text{a}$ is the frequency of the oscillating field,
and $\nu_0 = \gamma_{\text{H}} B_\text{0}$ is the proton Larmor frequency. 
In our case,
the factor is approximately $\xi \approx 0.02\,\text{cm}^{-3}$\,\cite{supp}.
The term $t e^{-t/T_2}$ describes the spin decoherence,
and $\text{sinc}(\Delta t/2)$ characterizes the effect of frequency detuning from the resonance.
We measure the transverse dipolar field $B_\text{s}$ generated by proton spins in acetonitrile molecules under a near-resonant oscillating field with an amplitude of $B_\text{a} \approx 0.6$\,nT.
The dipolar field is recorded using a $^{87}$Rb magnetometer, which is sensitive to both the transverse dipolar field and the applied oscillating field along $y$ [Fig.\ref{fig1}(a)].
The proton dipolar field is extracted from the raw data by removing the applied oscillating field [Fig.\,\ref{fig2}(a)].
In the near-resonance regime, where ${\text{sinc}(\Delta t/2) \approx 1}$,
the amplitude of the dipolar field reaches its maximum at approximately ${t \approx T_2 \approx 0.8\, \text{s}}$ before decaying.
The observed signal curve agrees well with the theoretical predictions in Eq.\,\ref{eq.BsoverBa}.

\begin{figure}[t]  
	\makeatletter
	\def\@captype{figure}
	\makeatother
	\includegraphics[scale=1.05]{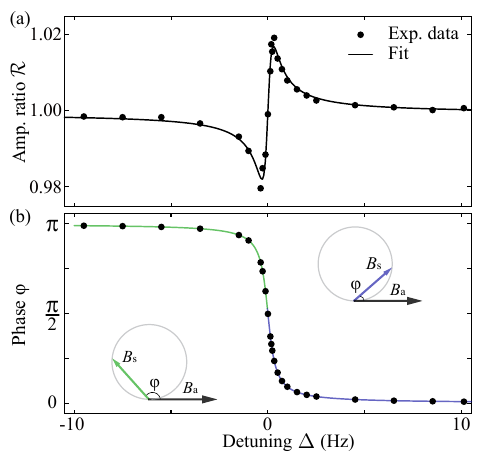}
	\caption{ 
    Anomalous frequency dependence of the proton amplification. (a) Dependence of the ratio $\mathcal{R}$ on the oscillating field frequency.
    (b) Phase difference $\varphi$ between the dipolar field and the oscillating field as a function of frequency.
    Colored lines indicate theoretical predictions.
    Insets illustrate the interference effect via vector diagrams of the dipolar field $B_{\text{s}}$ and the oscillating field $B_{\text{a}}$ in complex~plane.}
	\label{fig3}
\end{figure}

The amplification of hyperpolarized molecular nuclear spins is measured.
We acquire time-domain raw data that includes both the near-resonant oscillating field and the induced dipolar field.
In practical applications, extracting only the dipolar field from the raw data is impractical.
Alternatively, we perform a discrete Fourier transform on the acquired raw data, both with and without hyperpolarized spins, and then determine the amplification by comparing the spectral amplitudes at the frequency of the oscillating field.
As an example,
the experimental results for protons in acetonitrile are presented in the left panel of Fig.\,\ref{fig2}(b).
The spectral amplitude in the presence of hyperpolarized protons (denoted as Amp$_1$, blue curve) surpasses that from the oscillating field alone (denoted as Amp$_2$, black curve).
We calculate their ratio, $\mathcal{R} = \text{Amp}_1/\text{Amp}_2$,
and further show that this ratio is closely related to the factor $G$ by $\mathcal{R} \approx 1 + G \cos{\varphi}$\,\cite{supp}.
Here $\varphi$ denotes the phase difference between the dipolar field and the oscillating field,
with $\cos{\varphi} \approx 0.509$ in this case.
For acetonitrile,
$\mathcal{R}$ is measured to be $\mathcal{R} \approx 1.016$, 
corresponding to {$G \approx 3.1\%$}.
Similar experiments with pyridine yield ${\mathcal{R} \approx 1.012}$,
which corresponds to {$G \approx 2.4\%$}.
These results demonstrate the universality of amplification across various molecules.
Notably, the amplification factors observed here represent an order-of-magnitude enhancement over our estimations based on previous NMR spectroscopy data\,\cite{amp_NP2015, amp_NP2016PNAS, amp_NP2019, 2015SabrePy},
an improvement we attribute primarily to the higher number of polarized spins achieved in our experiments.
Moreover,
we show that the ratio $\mathcal{R}$, measured using acetonitrile, remains nearly constant as the oscillating field amplitude ranges from 0 to $0.9 \, \text{nT}$\,[Fig.\,\ref{fig2}(c)].
Beyond this range,
a gradual decrease in amplification is observed due to nuclear magnetization saturation\,\cite{2021NPaxion}.

\begin{figure}[t]  
	\makeatletter
	\def\@captype{figure}
	\makeatother
	\includegraphics[scale=1.04]{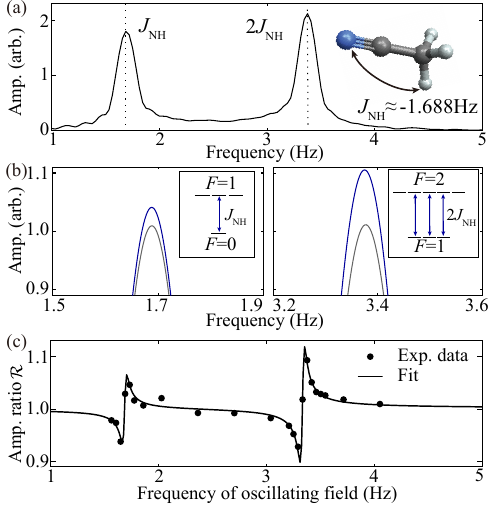}  %
	\caption{Magnetic amplification using hyperpolarized scalar-coupled spins. (a) Zero-field NMR spectrum of $^{15}\text{N}$-acetonitrile, with a scalar coupling \mbox{$J_{\text{NH}} \approx -1.688\,\text{Hz}$} between $^{15}\text{N}$ and $^1\text{H}$.
    (b) Amplification of the coupled spins near the $J_{\text{NH}}$ and $2J_{\text{NH}}$ frequencies.
    Colored curves show the spectral amplitude in the presence of hyperpolarized coupled spins, while gray curves correspond to the oscillating field alone. The corresponding transitions are shown in the inset.
    The complete spectra are provided in the Supplemental Material\,\cite{supp}.
    (c) Dependence of the ratio $\mathcal{R}$ on the oscillating field frequency.} 
	\label{fig4}
\end{figure}

We observe an anomalous amplification dependent on the field oscillation frequency.
A sequence of oscillating fields, each with a distinct frequency and an amplitude of approximately $0.6\,\text{nT}$,
is applied to hyperpolarized protons in acetonitrile.
Figure\,\ref{fig3}(a) shows the measured ratio $\mathcal{R}$ as a function of frequency detuning $\Delta$.
Notably,
$\mathcal{R}$ peaks at approximately 1.019 when $\Delta/2\pi \approx 0.34 \, \text{Hz}$,
rather than at $\Delta =0$.
Additionally, $\mathcal{R}<1$ is observed in certain detuning ranges, with a minimum of $\mathcal{R} \approx 0.979$ at $\Delta/2\pi \approx -0.35\,\text{Hz}$.
This dispersive frequency dependence
arises from interference between the dipolar field and the oscillating field\,\cite{supp}.
Both the dipolar field amplitude $B_\text{s}$ and the phase difference $\varphi$ between these fields, which are functions of frequency detuning,
dictate this interference.
Figure\,\ref{fig3}(b) shows $\varphi$ approaching 0 or $\pi$ with minimal sensitivity at large detunings,
while near resonance, $\varphi$ changes rapidly, reaching $\pi/2$ at $\Delta=0$.
To visually illustrate the interference, we represent the dipolar field and the oscillating field as vectors in the complex plane,
as depicted in the inset of Fig.\,\ref{fig3}(b).
For the dipolar field vector, the modulus $B_\text{s}$ represents its amplitude, and the angle $\varphi$ corresponds to the phase difference.
Under conditions where $B_\text{a} \gg B_\text{s}$, the
the ratio can be approximated as
${\mathcal{R} \approx 1 + G \cos{\varphi}}$\,\cite{supp}.
Constructive interference, occurring when $\varphi$ is between $0$ and $\pi/2$, leads to amplification,
with $\mathcal{R}$ peaking at $\varphi = \pi/4$, where $\mathcal{R} \approx 1 + \sqrt{2}G/2$.
Conversely, destructive interference occurs when $\varphi$ is between $\pi/2$ and $\pi$, resulting in $\mathcal{R}<1$.
The theoretical fitting curve agrees well with the experimental data~[Fig.\,\ref{fig3}(a)].

To further enhance amplification,
we extend our approach from a two-level system to a multi-level system governed by scalar spin-spin coupling, which remains largely unexplored as a potential physical sensor.
A key advantage of using such a scalar-coupled system is the presence of long-lived states\,\cite{2014longlivedPRL},
offering a promising path for enhanced amplification.
In our experiments, we use hyperpolarized $^{15}$N-labeled acetonitrile,
where the molecular spin system consists of one $^{15}$N nucleus and three equivalent $^{1}$H nuclei, forming an $^{15}\text{NH}_3$ spin cluster.
At zero magnetic field, spin dynamics are entirely governed by the scalar coupling Hamiltonian, {$H_{\text{J}} = 2 \pi \hbar \sum_{i=1}^{3} J_\text{NH} \mathbf{I}_\text{N} \cdot \mathbf{I}_\text{H}^{i}$},
which dictates that the energy eigenstates are also eigenstates of $F^2$ and $M_z$, where $F$ is the total angular momentum\,\cite{2011zfnmr, 2025zfnmr}.
For acetonitrile, the dominant coupling constant is $J_{\text{NH}} \approx -1.688\,\text{Hz}$.
Using the PHIP method, we achieve hyperpolarization of the $^{15}$N nucleus
and acquire the zero-field NMR spectrum shown in Fig.\,\ref{fig4}(a).
The spectrum exhibits two prominent peaks at frequencies of $J_\text{NH}$ and $2J_\text{NH}$,
corresponding to transitions within the $J_\text{NH}$ and $2J_\text{NH}$ manifolds of the scalar-coupled spins \cite{supp},
with signal-to-noise ratios (SNR) of approximately up to $1.72 \times 10^{3}$ and $1.97 \times 10^{3}$, respectively.
Notably, these manifolds exhibit remarkably long coherence times, with $T_2$ values of approximately 10.5\,s for the $J_{\text{NH}}$ manifold and 5.3\,s for the 2$J_{\text{NH}}$ manifold\,\cite{supp}, both values  exceeding the coherence times typical of two-level systems.

\begin{figure}[t]  
	\makeatletter
	\def\@captype{figure}
	\makeatother
	\includegraphics[scale=1.03]{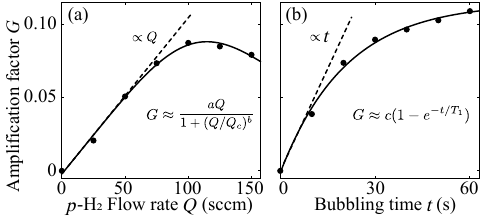}  %
	\caption{Dependence of the amplification on hyperpolarization parameters. The $2J_\text{NH}$ manifold of the scalar-coupled spins is used as an example. (a) Dependence of the amplification $G$ on $p$-$\text{H}_2$ flow rate $Q$.
    The experimental data (black dots) are fitted using an empirical formula with parameters ${a \approx 0.001}$, ${b \approx 4.0}$ and ${Q_\text{c} \approx 151.7\,\text{sccm}}$ (solid line), {where sccm denotes standard cubic centimeters per minute}.
    (b) Dependence of $G$ on $p$-$\text{H}_2$ bubbling time. The data are fitted with fit parameters $c \approx 0.11$ and $T_1 \approx 21.1\,\text{s}$.
    }
	\label{fig5}
\end{figure}


We measure the amplification of the hyperpolarized molecular scalar-coupled spins and observe significant responses at frequencies near both $J_\text{NH}$ and $2J_\text{NH}$ in Fig.\,\ref{fig4}(b).
The ratio reaches approximately $\mathcal{R} \approx 1.046$ near the $J_{\text{NH}}$ transition, corresponding to {$G \approx 6.5\%$},
and $\mathcal{R} \approx 1.093$ near the $2J_{\text{NH}}$ transition, corresponding to {$G \approx 13.2\%$}.
These scalar-coupled spins demonstrate a four-order-of-magnitude enhancement compared to Overhauser magnetometers\,\cite{2017Overhauser, 2025Overhauser, 2021Overhauser}, which have $G \approx 0.001\%$.
Similar to the two-level case depicted in Fig.\,\ref{fig3}(a),
we investigate the frequency dependence of the amplification in the scalar-coupled spins and observe two distinct peaks centered at $J_\text{NH}$ and $2J_\text{NH}$,
as illustrated in Fig.\,\ref{fig4}(c).
Near each transition frequency, the spin dynamics can be effectively simplified to a two-level system, with evolution governed by transitions within the respective manifold. A detailed theoretical justification for this effective two-level reduction is provided in the Supplemental Material\,\cite{supp}.

We further measure the dependence of the amplification on parameters in the hyperpolarization process,
which influences the initial polarization of molecular spins.
According to previous theories\,\cite{2016theorybarskiy,2016theoryknecht},
the molecular spin polarization is proportional to the concentration of $p$-H$_2$ in the liquid.
To maximize this concentration,
all our experiments are performed at a $p$-$\text{H}_2$ pressure of approximately $6\,\text{bar}$,
the maximum pressure safely supported by our setup.
Subsequently, we investigate the dependence of the amplification on $p$-$\text{H}_2$ flow rate $Q$ [Fig.\,\ref{fig5}(a)] and provide a potential explanation.
This dependence may be attributed to the variation in the total gas–liquid interfacial area, which influences the dissolution of parahydrogen into the liquid.
In the low-flow regime, we demonstrate that the gas-liquid interfacial area increases approximately linearly with $Q$\,\cite{supp},
resulting in $G \propto Q$ as shown in Fig.\,\ref{fig5}(a).
However, at higher flow rates, phenomena such as bubble coalescence and volume expansion limit further growth of the interfacial area\,\cite{supp},
causing $G$ to reach a maximum and subsequently decrease [Fig.\,\ref{fig5}(a)].
Overall, we propose a hypothetical model to capture the observed dependence, described by the empirical formula:
$G \approx a Q / \left(1 + (Q/Q_\text{c})^b \right)$ (see Supplemental Material for details\,\cite{supp}), where $Q_\text{c} \approx 151.7 \, \text{sccm}$ is the critical flow rate.
Although this model fits the experiment well, we emphasize that this process is not yet fully understood and requires deep investigation.
In addition, as PHIP is a dynamic process, the duration of gas bubbling $t$ is also an important parameter influencing the polarization.
During bubbling, polarization accumulates while simultaneously decaying due to spin-lattice relaxation, characterized by the time constant $T_1 \approx 21.1$\,s [Fig.\,\ref{fig5}(b)].
This continues until a steady state is established.
Our results can be well described by the theoretical formula $G \approx c(1 - e^{-t/T_1})$\,\cite{2016theoryknecht}.

Several promising applications exist for such molecular spin amplification,
even though the current amplification remains below unity.
Compared to proton and Overhauser magnetometers,
our approach offers several orders of magnitude higher amplification, making it particularly suitable for similar applications such as geomagnetic monitoring\,\cite{2017Overhauser}, mineral prospecting\,\cite{2016magnetometric}, and archaeological detection\,\cite{2018Overarchaeological}.
Additionally, the advanced capabilities of this approach open up new possibilities in cutting-edge scientific fields.
In absolute magnetometry, for example, the nuclei are effectively shielded by their electron shells,
providing exceptional immunity to environmental perturbations and enabling highly precise magnetic field measurements.
This unique feature offers a significant advantage in applications that require accurate determination of magnetic field intensity,
such as magnetic standards\,\cite{20203He} and spin-based gyroscope\,\cite{2005Gyroscope, 2021Gyroscope}.
In the realm of nuclear-dependent fundamental physics,
our method holds promise for the search for ultralight dark matter,
including axions and axion-like particles\,\cite{2014CASPEr, 2018axion, 2019natureaxion, 2021NPaxion, 2019axion},
which are well-motivated dark matter candidates.
These particles may couple with nuclear spins and behave as an oscillating pseudo-magnetic field,
an effect that can be enhanced using our nuclear spin sensor.

In conclusion, we have experimentally demonstrated magnetic amplification using hyperpolarized molecular nuclear spins, including protons and scalar-coupled spins.
Although our work is in its early stages, there is substantial potential for further enhancement.
For example, improvements in molecular spin polarization can be realized by
using deuterated catalysts to suppress relaxation\,\cite{amp_NP2016PNAS},
optimizing the polarization field sequence\,\cite{2024Multiaxisfields},
and selecting solvents with higher hydrogen solubility\,\cite{2025Solubility}. Additionally, nuclear-spin coherence time could be extended by an order of magnitude through the implementation of active feedback mechanisms\,\cite{2024Cooperative, 2025Joscillator, suefke2017hydrogen}.
With these advancements, we anticipate that the amplification factor could reach the order of $G \approx 100$,
unlocking significant potential for the development of ultrasensitive sensors.

This work was supported by the National Natural Science Foundation of China (Grants Nos.\,T2388102, 92476204, 12274395, 12261160569, and 12404341), the Innovation Program for Quantum Science and Technology (Grant No.\,2021ZD0303205), the Chinese Academy of Sciences (Grant No.\,XDB1300000), Youth Innovation Promotion Association (Grant No.\,2023474), and the New Cornerstone Science Foundation through the XPLORER PRIZE. This work was also supported by DFG/ANR grant BU 3035/24-1 and by the Alexander von Humboldt Foundation in the framework of the Sofja Kovalevskaja award to D.A.\,Barskiy.

\bibliographystyle{apsrev4-2}
\renewcommand{\refname}{References}
\bibliography{ref.bib}
\end{document}